\documentclass[conference]{IEEEtran}
\IEEEoverridecommandlockouts
\usepackage{cite}
\usepackage{hyperref}
\usepackage{amsmath,amssymb,amsfonts}
\usepackage{multirow}
\usepackage{multicol}
\usepackage{booktabs}
\usepackage{flushend}
\usepackage{algorithm} 
\usepackage{algpseudocode}

\usepackage{graphicx}
\usepackage{textcomp}
\usepackage{xcolor}
\usepackage{hyperref}
\usepackage[font=small,labelfont=bf,labelsep=space,figurename=Figure,skip=0pt]{caption}
\usepackage{float}
\usepackage{adjustbox}

\def\BibTeX{{\rm B\kern-.05em{\sc i\kern-.025em b}\kern-.08em
    T\kern-.1667em\lower.7ex\hbox{E}\kern-.125emX}}
\begin{document}

\title{Two-Phase Multi-Party Computation Enabled Privacy-Preserving Federated Learning \\
}
\author{%
  Renuga Kanagavelu\IEEEauthorrefmark{1},
  Zengxiang Li\IEEEauthorrefmark{1},
  Juniarto Samsudin\IEEEauthorrefmark{1},\\
  Yechao Yang\IEEEauthorrefmark{1},
  Feng Yang\IEEEauthorrefmark{1},
  Rick Siow Mong Goh\IEEEauthorrefmark{1},
  Mervyn Cheah\IEEEauthorrefmark{1}\\
  Praewpiraya Wiwatphonthana\IEEEauthorrefmark{1}\IEEEauthorrefmark{2},
  Khajonpong Akkarajitsakul\IEEEauthorrefmark{2}
  Shangguang Wang\IEEEauthorrefmark{3}\\
  \IEEEauthorblockA{%
  \IEEEauthorrefmark{1} \textit{Institute of High Performance Computing, A*STAR, Singapore}\\
 \IEEEauthorblockA{%
\IEEEauthorrefmark{2} \textit{King Mongkut's University of Technology Thonburi, Thailand}\\
\IEEEauthorrefmark{3} \textit{Beijing University of Posts and Telecommunications, Beijing, China}\\
}
}
}



\maketitle
\begin{abstract}
Countries across the globe have been pushing strict regulations on the protection of personal or private data collected. The traditional centralized machine learning method, where data is collected from end-users or IoT devices, so that it can discover insights behind real-world data, may not be feasible for many data-driven industry applications in light of such regulations. A new machine learning method, coined by Google as Federated Learning (FL) enables multiple participants to train a machine learning model collectively without directly exchanging data. However, recent studies have shown that there is still a possibility to exploit the shared models to extract personal or confidential data. In this paper, we propose to adopt  Multi- Party Computation (MPC) to achieve privacy-preserving model aggregation for FL. The MPC-enabled model aggregation in a peer-to-peer manner incurs high communication overhead with low scalability. To address this problem, the authors proposed to develop a two-phase mechanism by 1) electing a small committee and 2) providing MPC-enabled model aggregation service to a larger number of participants through the committee. The MPC- enabled FL framework has been integrated in an IoT platform for smart manufacturing. It enables a set of companies to train high quality models collectively by leveraging their complementary data-sets on their own premises, without compromising privacy, model accuracy vis-a`-vis traditional machine learning methods and execution efficiency in terms of communication cost and execution time.
\end{abstract}
\section{Introduction}
The astounding growth of data being collected and analyzed has led to rapid advances in data-driven technologies and applications~\cite{b1}. Every  data collected  is valuable as it contains insights towards various application domains.  With privacy laws being enforced by many countries globally over data collection of sensitive or personal data, this has a repercussion on how and what data is being collected.  The traditional way of just collecting data and processing it may no longer apply.  If systems were to collect just non-sensitive or non personal data, many important insights will be missed. Even if there is consent to collect sensitive or personal data, , data sharing without authorization would result in threats of data breaches ~\cite{b2}. Data collected could likewise be mishandled, thereby violating the fundamental rights to privacy for both individuals and companies~\cite{b3},~\cite{b4}.
The invention of advanced technologies to handle sensitive and personal information to maximise its value has become one of the biggest science growth for data analytics community in recent years. An interesting aspect that makes this growth
 even more important is that there are many scenarios that require the combination of obtaining sensitive information from several sources towards very useful insights. For example, hospitals, medical researchers and drug companies can benefit from jointly measuring the incidence of an illness without revealing private patient data; banks and financial regulators can assess systemic risk without revealing their private portfolios.However, in practice, data sharing across collaborators has various limitations including data ownership, legal compliance, and technical constraint~\cite{b5}. As a result, massive volume of data collected today remains in the form of fragmented data islands at individual organizations. 
\begin{figure}[!tbh]
\centering
  \includegraphics[width=0.45\textwidth]{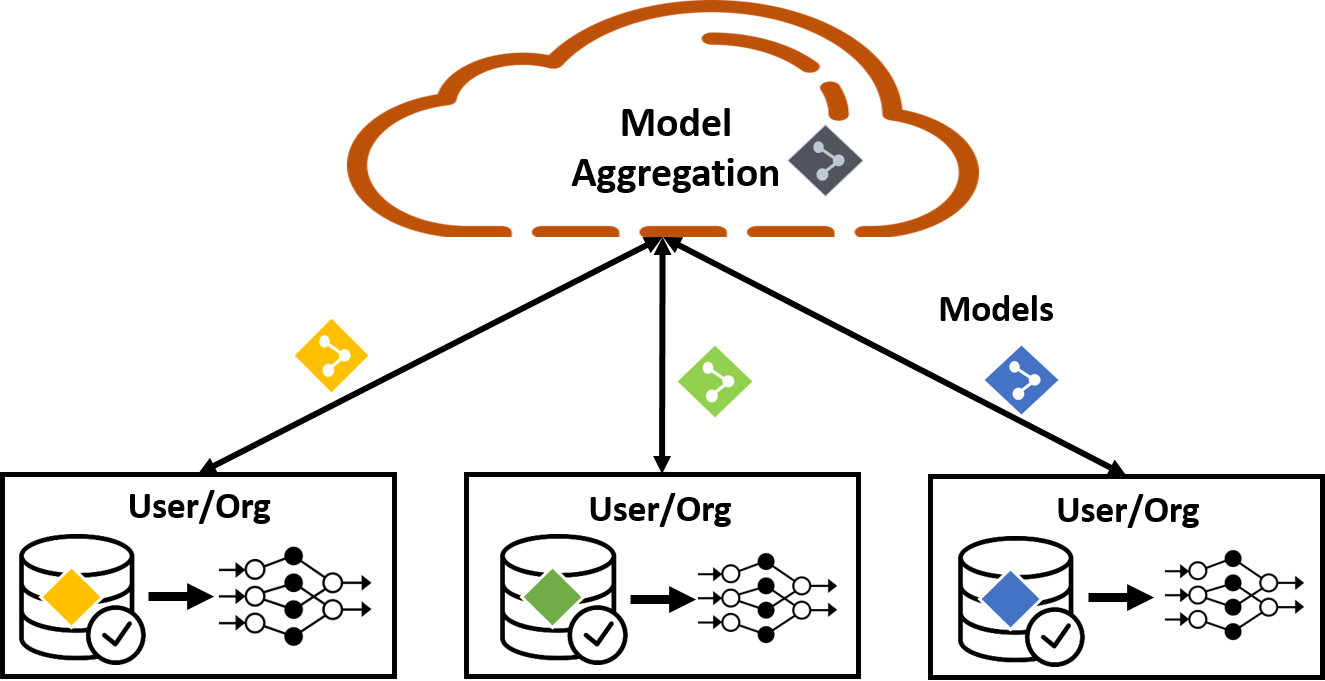}
\caption{Federated Learning Architecture}
\label{figure:FL}       
\end{figure}
Federated Learning (FL)~\cite{b6}, introduced by Google’s AI research team, supports decentralized collaborative machine learning over a number of devices or companies. The training data stays on each client device and only the locally trained model parameters are transferred to a central coordinator, which aggregate these models into a global federated model as shown in Figure~\ref{figure:FL}. The federated model is then sent back to the clients for training and improving the model iteratively. Google claimed that privacy is protected by keeping raw data on client devices and company’s on-premise IT infrastructure. In addition, federated learning could also make use of participants’ computation resources for training the time-consuming model.


Although the FL technique appears to ensure data remain on-premise, scientists have investigated that data (sensitive or otherwise) can still be extracted out from the FL technique  ~\cite{b7}.  They have published methods to show that the  machine learning model parameters can be    reverse  engineered  to extract the data sets that are residing with the devices or in the company’s IT infrastructure ~\cite{b8}, ~\cite{b9}.  Generally, it is concluded that in order to ensure the confidentiality of all the information and prevent the indirect data leakage, the model’s parameters need to be securely encrypted ~\cite{b10} against any possibility of reverse engineering and third party attacks against federated learning model training. For this reason, privacy-preserving techniques, such as secure Multi-Party Computation (MPC) ~\cite{b11}, Homomorphic Encryption (HE) ~\cite{b12}, ~\cite{b13} and Differential Privacy (DP)  ~\cite{b16}are typical technology candidates  to work together with FL for preserving the data confidentiality of model aggregation ~\cite{b9}. Google apparently recognized this flaw and subsequently  proposed a practical MPC-based secure aggregation for FL~\cite{b17} ; SecureBoost ~\cite{b18} which is implemented as a vertical tree-based FL system using homomorphic encryption to protect the gradients; and OpenMind ~\cite{b19} combined MPC and DP functionalities in its  FL framework.
Secure MPC is a class of cryptography techniques that allow a number of parties to compute a joint function over their sensitive data sets. Only the output of the joint function is disclosed without revealing participants' private inputs. Secure MPC performance~\cite{b15},~\cite{b16} has been improved significantly, due to the fast development of MPC protocols, software implementations, and underlying computation infrastructure. Today, secure MPC has known to be thousands of times faster than fully homomorphic encryption (FHE) implementation in typical applications. Another method to ensure data security and confidentiality is Differential Privacy (DP).  DP protects data privacy by adding random noise in the computation by a third party but it is not suitable for federated learning as the noise may affect the accuracy of federated learning model. For above reasons, this paper chooses MPC to support privacy-preserving (or data security and confidentiality of) model aggregation for federated learning.
However, there is a downside if one were to employ MPC to preserve the data confidentiality of FL.  In general, MPC protocols require all parties to generate and exchange secret shares of private data to all other parties. This basic needs inevitably results in high communication overhead and this overhead exponentially increases with the number of parties in the membership list agreeing to work together, regardless if they are trusted or non trusted parties. This leads to the motivation of this paper and the technique the authors are publishing.  To reduce the communication cost and improve scalability, this paper proposes a two-phase MPC-enabled FL framework. What it means is that instead of a condition that every member is required to generate and exchange secret shares of data across all members list, it proceed to elect a subset of FL members as the model aggregation committee members out of the whole membership list.  The elected committee membersand then uses MPC service to aggregate the local models of all FL parties.  The 2-phase MPC introduces a hierarchical structure, where there is a need for fewer secret shares being exchanged for privacy-preserving (or data confidentiality) model aggregation. The technique becomes more pronounced when the membership lists is large, and the number of committee members is only a fraction of the number of FL parties.
Both Peer-to-Peer and Two-Phase MPC-enabled FL frame- work are implemented on PyTorch, a Facebook-based  distributed machine learning and deep learning platforms. A parallel mechanism is developed to enable MPC-based model aggregation on the entire model tensors with large amounts of parameters. The FL framework is further integrated with our in-house Industrial IoT (IIoT) platform which could be deployed on-premise or on public Cloud. A smart manufacturing use case with real-world sensor data sets and machine failure labels is used in our experiments, to demonstrate the effectiveness of the proposed FL framework in terms of model accuracy and execution efficiency.
%
\section{Related Work}
Federated Learning (FL) is too broad a subject.  Thus, FL can be further categorized into horizontal FL and vertical FL. In horizontal FL, all participating parties (e.g. regional hospitals or manufacturing companies using the same type of machine)have the entire feature set and labels available to train their local model. The computed local gradients from different participants are being sent (to where is it being sent?) to the server/aggregator to train a global model. In vertical FL, all participating parties (e.g. banks, insurance and e-commerce companies within the same city) collect data with different feature spaces and only one party has access to the label. Parties (without label) are unable to train a model on their own, due to the lack of information. As a result of this, additional layer of complexity is added to securely align the sample spaces and the training process requires exchanging partial model updates. In this paper, the authors focus on horizontal FL only.



Currently, only a few production-grade alternatives that challenge the centralized machine learning paradigm are available. Google has started to work towards Federated Learning called Tensor flow Federated (TFF)~\cite{b22} in Mar 2019. CrypTen ~\cite{b44} is a MPC-based privacy preserving machine learning framework built on PyTorch by Facebook. OpenMined ~\cite{b24}, an open-source community focused on building technology that combines Deep Learning, Federated Learning, Homomorphic Encryption and Blockchain over decentralized data. FATE (Federated AI Technology Enabler)~\cite{b25} is another open-source project initiated by WeBank’s AI Group to provide a secure computing framework for building the federated AI ecosystem. ByteLake~\cite{b26}, an AI consultancy based in Poland, recently released a proof of concept for the manufacturing industry in concert with Lenovo for predictive maintenance. Different from existing general FL frameworks, our federated learning framework supports privacy-preserving, value oriented federated learning over decentralized data that is integrated into IIoT platform with visualization and smart manufacturing domain knowledge. 
Although MPC has been widely applied for privacy-preserving FL, limited work has been published to enhance performance and scalability of MPC-enable FL frameworks. OpenMined~\cite{b24} provides a very basic evaluation of MPC overhead for model aggregation. Google provides complexity analysis and communication overhead evaluation on its practical MPC protocols for FL. In contrast, this paper proposes a hierarchical and parallel MPC operations for efficient model aggregation.  
\section{MPC-enabled Federated Learning Framework}
\label{MPCFLFramework}
In this section, the architecture and implementation details are illustrated for both traditional Peer-to-Peer and  Two-Phase MPC-enabled federated learning framework. In addition, theoretical analysis on the number and size of messages exchanged is also provided. Some variables are defined and self-explained in Table~\ref{FLVariables}.
\begin{table}[t]
\scriptsize
\renewcommand{\arraystretch}{1.3}
\centering

\caption{Variables in Federated Learning Framework}\label{FLVariables}{
\centering
\begin{tabular}{|p{0.65in}|p{0.35in}|p{0.3in}|p{1.3in}|}
\hline \textbf{Category} & \textbf{Symbol} &  \textbf{Type} & \textbf{Description}  \\
\hline\multirow{4}{*}{Federated Learning Framework} & $n$ & Int & Number of parties \\
\cline{2-4} & $t$ & Int & Number of iterations in local model training\\
\cline{2-4} & $e$ & Int & Number of epochs in global FL model training\\
\hline \multirow {4}{*}{Model Aggregation Committee} & $m$ & Int & Number of elected committee members \\
\cline{2-4} & $b$ & Int & Batch size of each round of committee election \\
\cline{2-4} & $C$ & Array & List of committee members\\
\hline \multirow{7}{*}{Neural Network Model} & $T(i, j, k)$ & Tensor & Parameters/weights of local model of $i^{th}$ party at $j^{th}$ global epoch and $k^{th}$ local iteration \\
\cline{2-4} & $G(j)$ & Tensor & Parameters/weights of aggregated model of $j^{th}$ epoch \\
\cline{2-4} & $s$ & Integer & The size of parameters/weights of local/aggregated models\\
\hline
\end{tabular}
\vspace{-2ex}
}
\end{table}
\subsection{Multi-Party Computation}
\label{MPCProtocols}
Secure MPC is a class of cryptographic techniques that allow for confidential computation over sensitive data. The two dominated MPC techniques \cite{b20} today are garbled circuits and secret sharing. Garbled circuit is a cryptographic protocol based on Boolean circuit. It was proposed to solve the popular millionaire problem, describing two millionaires want to know who is richer without revealing their actual wealth. 
\begin{figure}[!tbh]
\centering
  \includegraphics[width=0.45\textwidth]{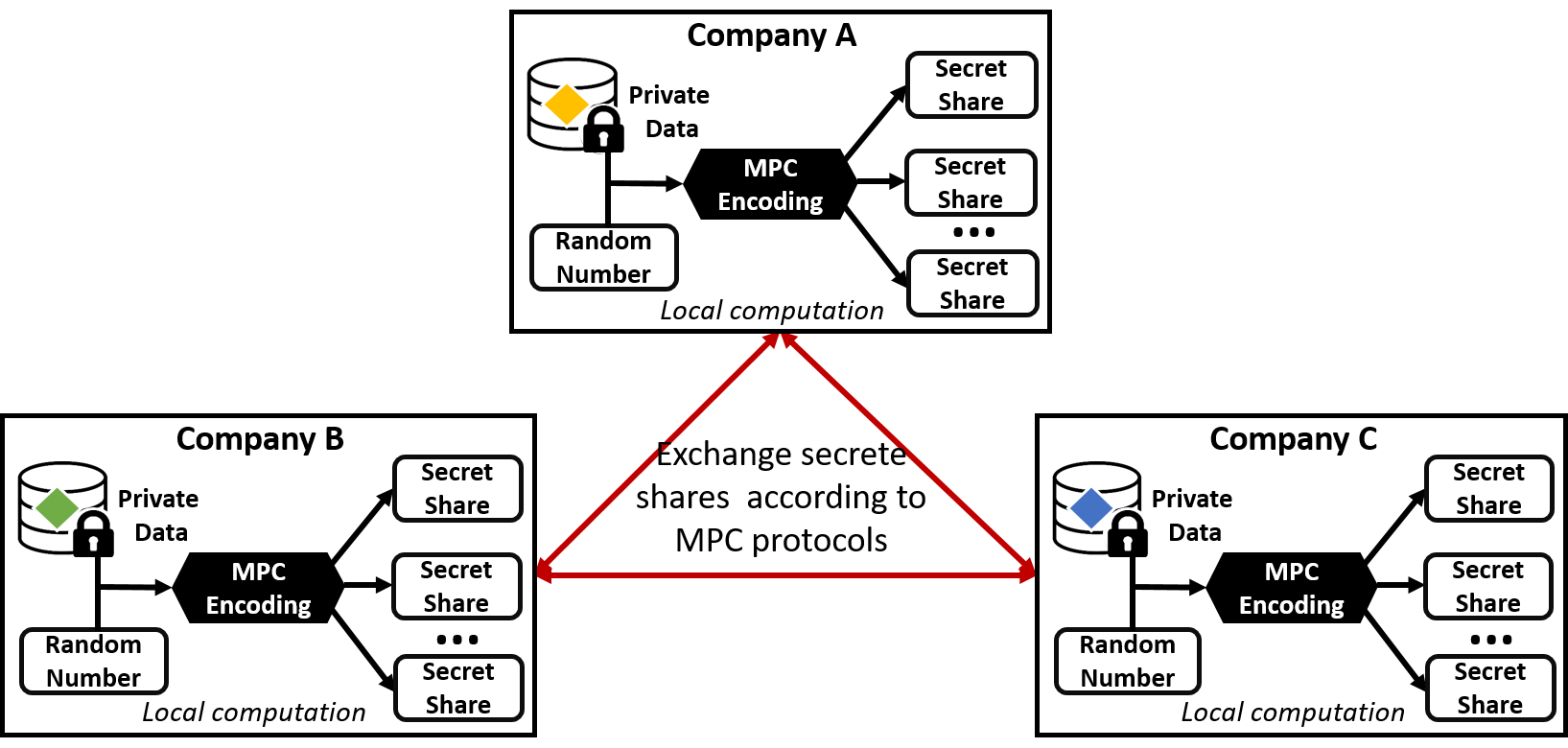}
\caption{Secure Multi-Party Computation}
\label{fig:MPC}       
\end{figure}
Recently, secret sharing based MPC protocols have been more commonly used in production systems~\cite{b21}. A typical secret share MPC protocol is shown in Figure~\ref{fig:MPC} , each sensitive data is split into “secret shares”, which in combination yield the original data. They interact with each other to compute confidential function by exchanging secret shares according to MPC protocols. As long as majority of parties honestly follow the protocol, no party learns anything beyond the final output of the computation. Since each computing party works on one piece of secret share, only if a threshold number of computing parities are compromised, the attacker could reconstruct private data from secret shares. Consequently, MPC can be viewed as a cryptography method for providing the functionality of a trusted party—who would accept private inputs, compute a function and return the result to the stakeholders—without the need for mutual trust among participants. 
Both Additive~\cite{b28} and Shamir~\cite{b39} secret sharing MPC protocols are used for privacy-preserving model aggregations in FL framework. For simplicity consideration, this study focus on Neural Network (NN) machine learning models only and the model aggregation calculates the averaged weights among local NN models trained by all parties. Since the number of FL parties are pre-known, only addition MPC operation is needed.
\begin{algorithm}
\small
	\caption{Addition operation using Additive secret sharing MPC protocol}
	\label{additionAdditive}
	\begin{algorithmic}[1]
		\State{Function\{\textbf{Additive.add(V, n)\}}
		\State{$Q = a\ very\ large\ prime\ number$}
		\For{$i\in[1,n]$} 
		\State{\#All parties do this loop concurrently}
		\For{$j \in [1, n-1]$}
		\State{$V(i,j) = (Random\ Number)$}
        \EndFor
        \State{$V(i,n) = (V(i) - \sum_{j=1}^{n-1} V(i,j)) \mod Q$}
		\For{$j \in [1, n] \& j\neq i$}
	        \State{Send $V(i, j)$ to\ $j^{th}$\ party}
		    \State{Receive $V(j, i)$ from $j^{th}$ party}
		\EndFor
		\State{$S(i) = \sum_{j=1}^{n} V(j,i)$}\\
		\EndFor
		\For{$j \in [1, n] \& j\neq i$}
	        \State{Send $S(i)$ to\ $j^{th}$\ party}
		    \State{Receive $S(j)$ from $j^{th}$ party}
		\EndFor
		\State{$S = \sum_{i=1}^{n} S(i)$}\\
		\Return{$S$}
		\State{EndFunction}}
	\end{algorithmic} 
\end{algorithm}
The addition operation (i.e., the sum of private value $V(i)$ among $n$ parties) using Additive secret sharing MPC protocol is shown in Algorithm~\ref{additionAdditive}. Each party generates $n-1$ secret shares for its private value using random numbers. The last piece of secret share is calculated as $V(i,n) = (V(i) - \sum_{j=1}^{n-1} V(i,j)) \mod Q$. The meaningless secret shares are exchanged with other parties and then each party calculate the sum of one piece of secret share of the private values from all parties. The sum value calculated at each party is further exchanged and accumulated to get the addition result $S$. By adopting Additive MPC protocol, all parties conduct confidential computation on meaningless secret shares. The introduced randomness could be eliminated by two rounds of message exchange and addition, and it is straight forward to prove that $S = (\sum_{i=1}^{n} V(i)) \mod Q$. The confidential computation only discloses the final output $S$. It is impossible to deduce parities' private value $V$, if the parties are honest-but-curious without collusion.
On contrary, Shamir secret sharing protocol uses polynomial interpolation and is secure under a finite field. In order to generate secret share $v$, it uses a random $d$ degree polynomial $q(x)=a_0 + a_1 \times x + a_2 \times x^2 + a_d \times x^d$, where $a_0$ = $r$, and $\forall i \in [1, t]$ $a_i$ are random numbers. Given any $d+1$ shares, polynomial $q(x)$ could be reconstructed using Lagrange interpolation. Hence, the secret $v$ could be trivially recovered using $v = q(0)$. We choose $d=n-1$, that is all parties generate $n$ secret shares for each private data. The secret shares are exchanged with other parties for the addition operation on their private data without compromising the privacy. The security level and computation cost of Shamir secret sharing protocol are much higher than Additive  protocols. However, the number and size of messages exchanged remains the same. 
\subsection{Peer-to-Peer MPC-enabled Federated Learning}
\begin{figure}[!tbh]
\centering
  \includegraphics[width=0.48\textwidth]{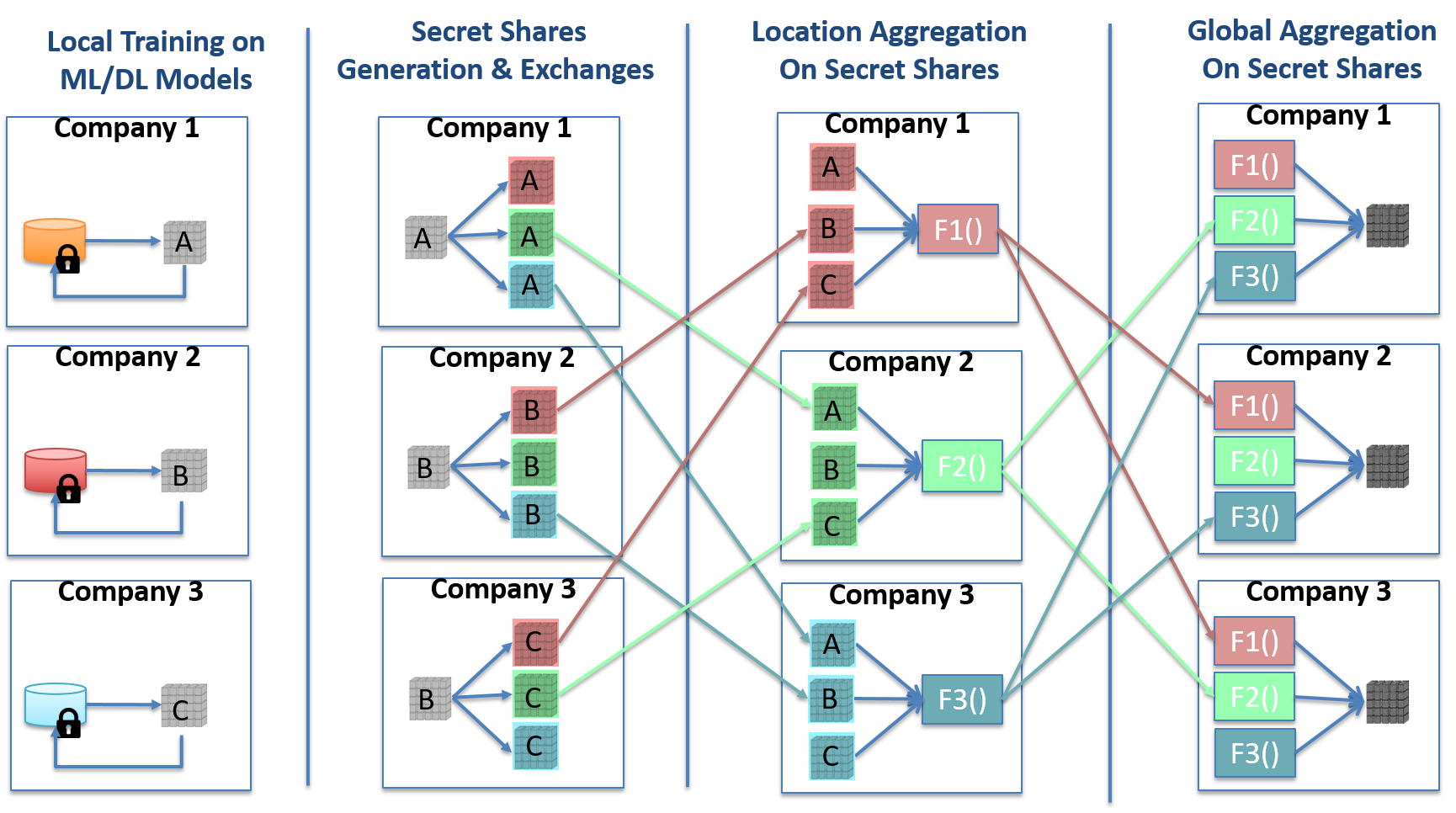}
\caption{MPC-enabled Privacy-Preserving Model Aggregation }
\label{figure:powermpc}       
\end{figure}

We consider the standard settings for federated learning, where participants concurrently train their local NN models on their private data sets. The objective is to find a tensor $T$ of parameters/weights of the NN model that minimizes the empirical loss function. To enhance the privacy without sacrificing accuracy, the tensor $T$ is securely aggregated using MPC technique. Once the convergence criterion is met, the training will be completed. 
As illustrated in Section~\ref{MPCProtocols}, MPC introduces overhead on both computation and communication. Hence, the performance enhancement is essential to make privacy-preserving FL practicable. The machine learning models become more and more complex to solve large-scale and complex problems. Sequential MPC operations on individual parameters/weights of the model would introduce tremendous overhead. The efficiency of parallel MPC operations has been verified in~\cite{b40}, as the computation and commutation cost could be reduced significantly by doing so in bulks. 
Figure~\ref{figure:powermpc} illustrates the procedure of model aggregation by calculating the average value of the tensors. The tensors of individual local models are privacy-sensitive. They are split into multiple tensors as the secret shares by adding randomness. After exchanging secret shares, each participant holds one share of each tensor. Consequently, participants conduct a local aggregation of the secret shares, followed by a global aggregation to cancel out the randomness and thus get the average value of the tensors (i.e., local models) of all participants. In this way, participants get federated model for the global training epoch, while keeping their data and local models in private.   
The number of messages denoted as $Msg\_Num$ exchanged in the model aggregation can be calculated as follows 
\begin{equation}\label{msgNumPeer}
\small
  Msg\_Num= (n\times (n-1)) \times 2 \times e =  2 \times n^2 \times e  - 2 \times n \times e  
\end{equation}
\noindent The size of messages denoted as $Msg\_Size$ exchanged in the model aggregation is
\begin{equation}\label{msgSizePeer}
\small
  Msg\_Size= Msg\_Num \times s =  2 \times n^2 \times e \times s  - 2 \times n \times e \times s  
\end{equation}
\noindent 
Since the computation complexity of the above two Equations is  $O(n^2)$, the traditional Peer-to-Peer MPC-enabled FL framework has low scalability~\cite{b30}. On the other hand, in practice, when more and more companies join the federate learning ecosystem, the  benefit achieved  becomes more and more significant. 
\subsection{Two-Phase MPC-enabled Federated Learning}
\begin{figure}[!tbh]
\centering
  \includegraphics[width=0.48\textwidth]{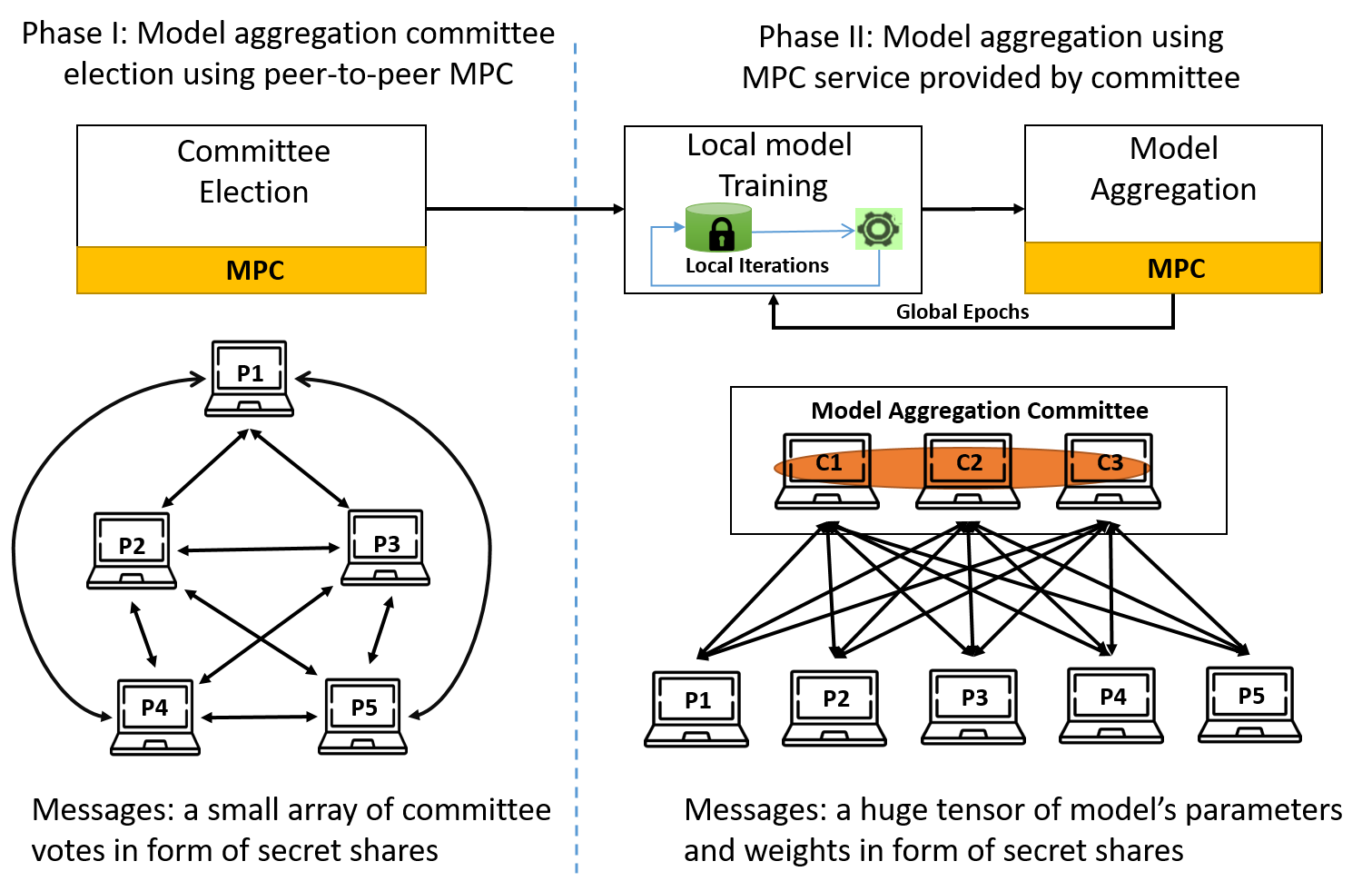}
\caption{Two-phase MPC-enabled federated learning framework}
\label{figure:elempc}       
\end{figure}

In order to improve the scalability, we propose a two-phase MPC-enabled FL framework, which avoids exchanging large amounts of huge tensor of model's parameters/weights (in the form of secret shares) across all FL parties. As shown in Figure~\ref{figure:elempc}, $Phase\ I$ uses peer-to-peer MPC to elect a subset of FL parties as the model aggregation committee members. $Phase\ II$ uses the MPC service provided by committee members to aggregate the local models of all FL parties. 
\begin{algorithm}
\small
	\caption{Phase I: Committee Election}
	\label{committeeElection}
	\begin{algorithmic}[1]
		\State{Function\{\textbf{Committee.election(n, k, b)\}}
		\For{$i\in[1,n]$} 
		\State{\#All FL parties do this loop concurrently}
		\While{$C.length < m$}
		\For{$w \in [1, b]$}
		\State{$B(i).append (a\ random\ number \in [1, n])$}
        \EndFor
		\For{$j \in [1, n]$}
		    \State{Generate\ secret\ share\ $B(i, j)$ for $B(i)$}
		\EndFor
		\For{$j \in [1, n] \& j\neq i$}
	        \State{Send $B(i, j)$ to\ $j^{th}$\ party}
		    \State{Receive $B(j, i)$ from $j^{th}$ party}
		\EndFor
		\State{$B(i) = \sum_{j=1}^{n} B(j,i)$}\\
		\For{$j \in [1, n] \& j\neq i$}
	        \State{Send $B(i)$ to\ $j^{th}$\ party}
		    \State{Receive $B(j)$ from $j^{th}$ party}
		\EndFor
		\State{$B = \sum_{i=1}^{n} B(i)$}\\
		\State{$B$ = $B$ $\mod$ $n$}
		\For{$part\ index\ in\ B$}
		\State{$C.append(highest\ vote\ party\ in\ B)$}
		\EndFor
		\EndWhile
		\EndFor
		\Return{$C$}
		\State{EndFunction}}
	\end{algorithmic} 
\end{algorithm}
The pseudo code of $Phase\ I$  committee election function is shown in Algorithm~\ref{committeeElection}. It selects $m$ out of $n$ FL parties to establish a model aggregation committee. After all parties execute the function concurrently, they would get the list of committee members (i.e., $C$). All parties generate a batch of random numbers as their initial votes. The votes of all parties are added together through the MPC protocol to obtain aggregated votes, without revealing the initial votes from all parties. 
The cost of $Phase\ I$ is marginal, as it is executed once only for each FL training and the message exchanged is a small array of committee votes (in form of secret shares). The number of messages denoted as $Msg\_Num$ exchanged in $Phase\ I$ is 
\begin{equation}\label{msgNumPhaseI}
\small
  Msg\_Num= (n\times (n-1)) \times 2  =  2 \times n^2   - 2 \times n   
\end{equation}
\noindent The size of messages denoted as $Msg\_Size$ exchanged in $Phase\ I$ is
\begin{equation}\label{msgSizePhaseI}
\small
  Msg\_Size= Msg\_Num \times b =  2 \times n^2  \times b  - 2 \times n  \times b  
\end{equation}
\noindent 
\begin{algorithm}
\small
	\caption{Phase II: Model Aggregation}
	\label{modelAggregation}
	\begin{algorithmic}[1]
		\State{Function\{\textbf{Model.aggregation(n, e, t, m)\}}
		\For{$j \in [1, e]$}
		    \For{$i \in [1, n]$}
		    \State{\#All FL parties do this loop concurrently}
		        \For{$k \in [1, t]$}
		            \State{T(i, j, k) = trained local model}
		        \EndFor
		        \For{$w \in [1, m]$}
		            \State{Generate\ secret\ share\ $T(i, j, t, w)$ for $T(i, j, t)$}
                    \State{Upload $T(i, j, t, w)$ to $w^{th}$ committee\ member}
		        \EndFor
		    \EndFor   
		    \For{$w \in [1, m]$}
		    \State{\#All committee members do this loop concurrently}
		    \State{$G(w,j)=\sum_{i=1}^{n} T(i, j, t, w)$}
		    		\For{$v \in [1, m] \& v\neq w$}
	                    \State{Send $G(w,j)$ to\ $v^{th}$\ party}
		                \State{Receive $G(v, j)$ from $v^{th}$ party}
		              \EndFor
		    \State{$G(j)=\sum_{v=1}^{m}G(v,j)$}
		    \For{$i \in [1, n]$}
		    \If{$i\mod m = w-1$ }
		    \State{Send $G(j)$ to $i^{th}$ party}
		    \EndIf{}
		    \EndFor
		\EndFor
		\EndFor
		\Return{$G(e)$}
		\State{EndFunction}}
	\end{algorithmic} 
\end{algorithm}
The pseudo code of $Phase\ II$  model aggregation function is shown in Algorithm~\ref{modelAggregation}. All FL parties trained models locally with $t$ iteration and get local model $T(i, j, t)$ (Line 5 to 7). They generate and upload $m$ secret shares of their local models to corresponding committee members (Line 8 to 11 ). Thereafter, committee members work together to conduct confidential model aggregation (Line 15 to 20), and then broadcast the aggregated model of this epoch $G(j)$ to all FL parties. FL training and model aggregation is conducted in pre-known number of global epochs or until the model is converged. 
The cost of $Phase\ II$ is controlled by reducing the number of secret shares of huge tensors from $n$ to $m$, where  $m<<n$. The number of messages denoted as $Msg\_Num$ exchanged in $Phase\ II$ can be calculated by adding following items 1) all parties upload secret shares to committee; 2) Committee members do model aggregation; and 3) committee members broadcast the aggregated model to all parties.		
\begin{equation}\label{msgNumPhaseII}
\small
  Msg\_Num= (n \times m   +  (m-1)  + n) \times e   =  (n \times m  + n +  m-1)  \times e   
\end{equation}
\noindent The size of messages denoted as $Msg\_Size$ exchanged in $Phase\ II$ is
\begin{equation}\label{msgSizePhaseII}
\small
  Msg\_Size= Msg\_Num \times s =   (n \times m  + n +  m-1)  \times e  \times s
\end{equation}
\noindent 
By accumulate $Msg\_Num$ and $Msg\_Size$ of both $Phase\ I$ and $Phase\ II$,  the communication cost of the proposed two-phase MPC-enabled FL framework is illustrated in following equations. 
\begin{equation}\label{msgNumTwoPhase}
\small
  Msg\_Num=    2 \times n^2    +  n \times (m \times e  + e  -2 ) + m \times e  - e 
\end{equation}
\begin{equation}\label{msgSizeTPhase}
\small
Msg\_Size= 2 \times n^2 \times b   +  n \times (m \times e \times s + e \times s -2 \times b) + m \times e \times s - e \times s 
\end{equation}
\noindent By comparing with traditional Peer-to-Peer, the proposed Two-Phase MPC-enabled FL framework could significantly reduce both the number and size of messages exchanged. The scalability is also improved significantly, by avoiding generating $n$ secret shares of huge tensors (in form of secret shares) and exchanges them with all parties. Experimental results for a particular use case and parameter configuration are further illustrated in Section~\ref{CommunicationResults}.  
\subsection{Integrated Federated Learning with IIoT Platform for Smart Manufacturing}

Recent advancement in IoT and AI have push fast transformation in smart manufacturing. Industrial IoT (IIoT) platforms like Azure, AWS, ThingWorx and MindSphere support and orchestrate data collection, storage, processing and visualization on edge, on-premise and Cloud. Predictive maintenance~\cite{b19} has become a very strong use case for manufacturers advancing to Industry 4.0. It offers an effective solution for detecting faults (diagnosis) and predictions of the future working conditions and the remaining useful life (prognosis), using the data captured from IoT devices (e.g., vibration, force, acoustic and emission sensors). This helps to avoid system interruptions and thereby minimizing the downtime and cost.
In order to provide high quality predictive maintenance service, the equipment vendor requires customers to share data, including equipment operation conditions, sensor data and failure incidences. By aggregating the data sets from a number of customers, the vendor could train and update the predictive maintenance which taking different operation scenarios into considerations. However, this centralized training mechanism faces lots of challenges e.g., data ownership, data privacy, latency and cost.

\begin{figure}[!tbh]
\centering
  \includegraphics[width=0.48\textwidth,height=0.2\textheight]{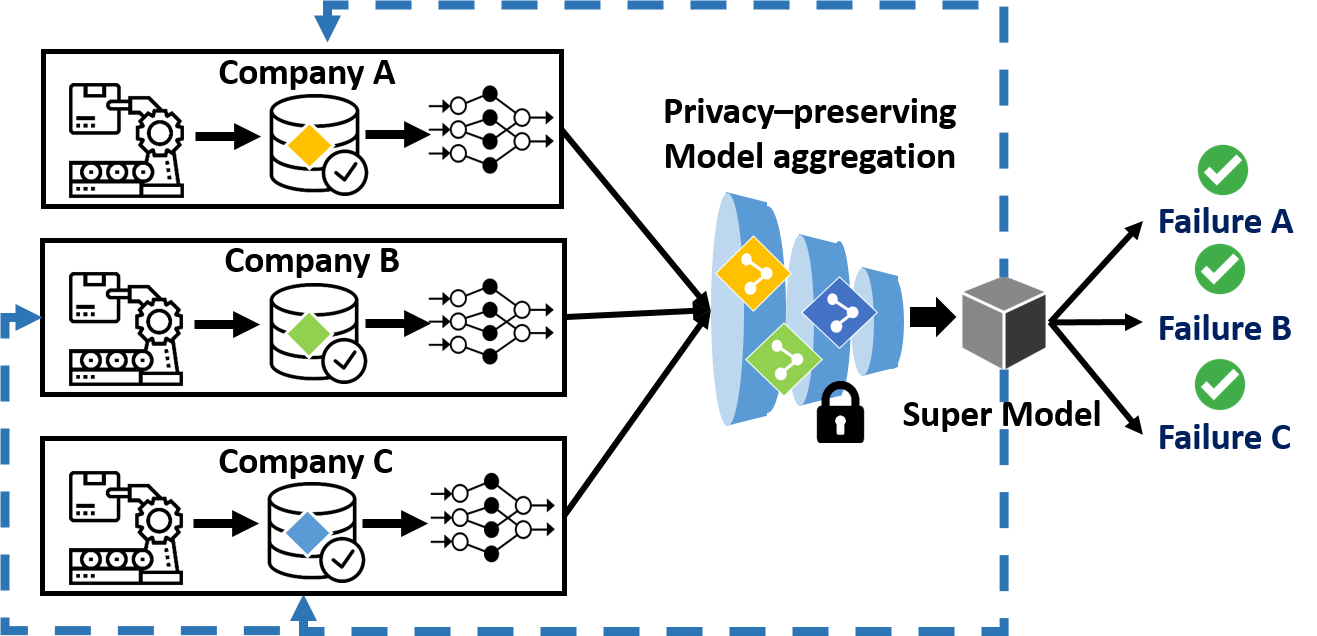}
\caption{ Federated Learning for Smart Manufacturing}
\label{figure:FL_IIOT}       
\end{figure}

Federated learning as shown in Figure~\ref{figure:FL_IIOT} could address the issues of centralized training mechanism. It has following features:
\begin{itemize}
\item Sensor/feature data are always kept at on-premise;
\item Companies train models locally and concurrently;
\item Local models are aggregated using a secure MPC protocol in a privacy-preserving manner; 
\item Models are trained and aggregated iteratively to continuously improve the accuracy; and
\item Federated model is able to predict failures under different operation conditions.
\end{itemize}

\begin{figure}[!tbh]
\centering
  \includegraphics[width=0.48\textwidth,height=0.2\textheight]{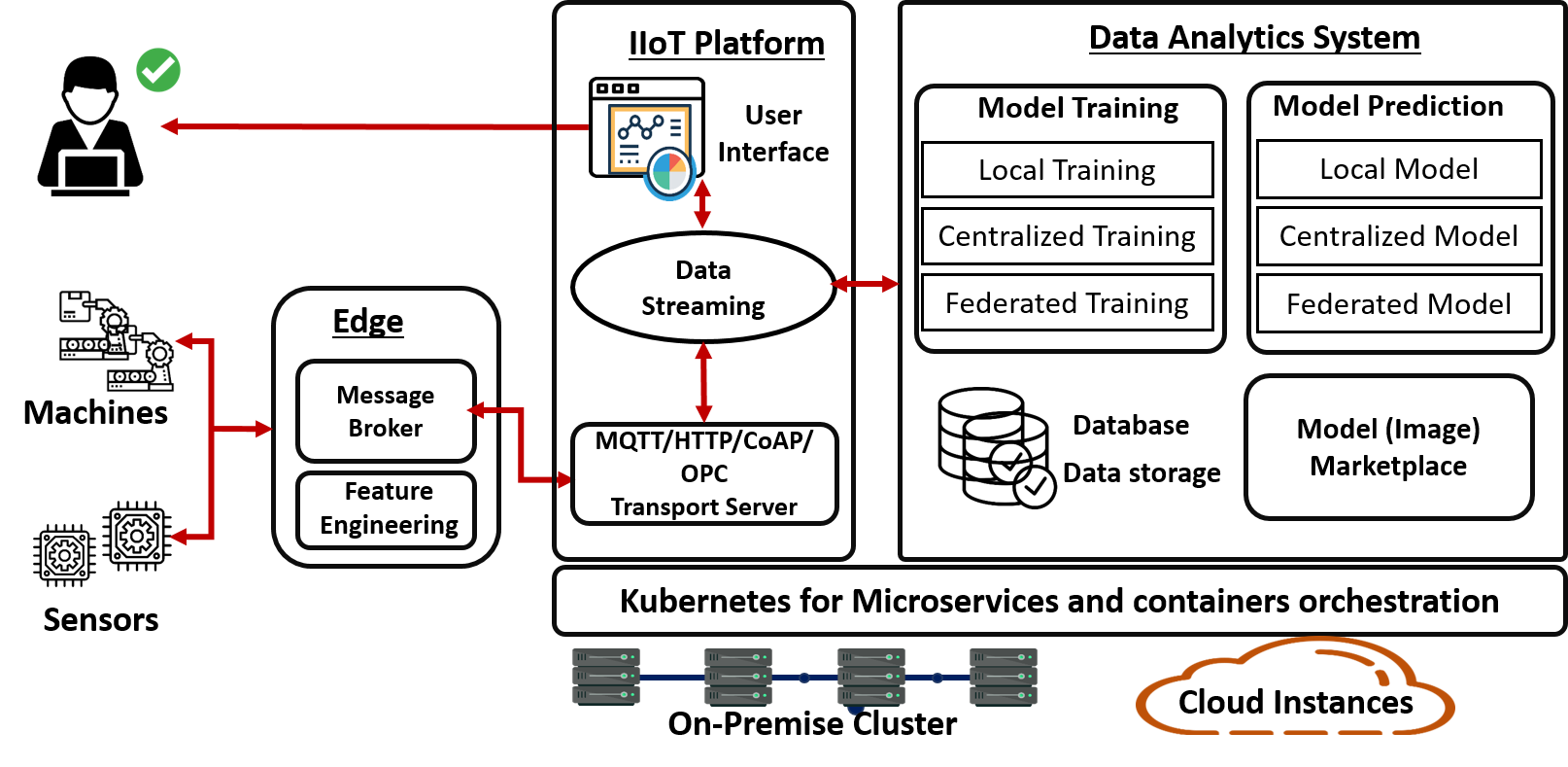}
\caption{ Federated Learning Enabled IIoT Platform}
\label{figure:integrate}       
\end{figure}
The proposed federate learning framework is integrated with our in-house IIoT platform for smart manufacturing applications as shown in Figure~\ref{figure:integrate}. The integrated system includes following components:
\begin{itemize}
\item  \textbf{Edge device} -  which can be a MiniPC or Raspberry PI, collects data from machine or sensors, and extracting features from the raw data. Consequently, the feature data are pushing to the IIoT platform through standard communication protocols, e.g., MQTT and OPC UA. 
\item  \textbf{IIoT platform}- can be deployed on on-premise or cloud servers. It receives streaming data from edge device and then visualize it on dashboard for real-time monitoring.
 \item  \textbf{Data analytics system} - can also be deployed on on-premise or cloud servers. The streaming data are saved into a database. Using the historical data, companies could train machine learning models in different manners (i.e., Locally, Centralized and Federated).  These models with different versions are packaged as docker containers and stored in model marketplace. Companies could deploy these models to provide predictions using real-time streaming data. The prediction and insights are helpful for operators to improve manufacturing process and efficiency.  
\end{itemize}
\section{Experiment Design and Results}
\subsection{Use case: Fault Detection in Electrical Machines}
\label{usecaseSection}

In order to evaluate the proposed federated learning framework, this paper studied a use case of fault detection in electrical machines which are widely adopted in smart manufacturing. Effective diagnosis of motor faults is essential to enhance reliability of manufacturing process as well as to reduce the cost of operation and maintenance. This study used the real world sensor data (e.g., vibration, temperature, currency and acoustic) and machine healthy and faulty label from a number of induction motors. The data sets were collected in the University of Tennessee USA, by Prof. J. Wesley Hines’s team, where each motor was subject to thermal aging processes \cite{b36}. In the same way as conducted in \cite{b37}, the sensory data were pre-processed and the time-domain features were extracted, with the top 121 features being employed here in this study. Due to the different total lifetimes, it can be reasonably assumed that the health status of motors at each life cycles are different, mimicking the different fault types of different companies.
We choose PyTorch 1.2.0 and Python 3.73 as machine learning library in our federated learning framework for following reasons: easy to use API, multi GPU support, python support, custom data loaders and simplified pre-processors. In our experiment, we made the assumption that all companies are using the same features extracted from sensor data. They are using the Neural Network (NN) models provided by PyTorch for fault detection. Both simple and complex neural network architecture structures were used. In the simple NN model, it only has input (121 features) and output layers (2 prediction results). That is the size of simpleNN model (i.e., $s$ in table~\ref{FLVariables}) is equal to 242. In the complexNN model, another hidden layer with 60 neurons is added. Hence, the size of complexNN model is equal to 7380.   
\subsection{Experimental Design}
In our experiments, we evaluated the performance of proposed federated learning platform using both Additive and Shamir secret sharing MPC protocols. Experiments are conducted to compare traditional peer-to-peer and proposed two-phase federated learning in terms of messages exchanged and execution time. Experiments are conducted on following three different environments:
 \begin{itemize}
\item Local-SingleServer: All parties run corresponding processes on a single local server with an Intel(R) Xeon(R) CPU i7-7700 V8 (3.60GHz) and 32GB of RAM, and Ubuntu 18.04 OS. 
\item AWS-SameRegion: All parties run on 16 EC2 virtual machine (VM) instances at the same region (i.e., Singapore)
\item AWS-CrossRegion: All parties run on 16 EC2 virtual machine (VM) instances at different eight regions (i.e., Singapore, Canada, London, Ireland, Tokyo, Ohio, Sao Paulo and Mumbai)
\end{itemize}
AWS EC2 VMs are t3.medium instances, each of which has 2 vCPUs and 4 GiB RAM, and Ubuntu 18.04 OS.
\subsection{Model Prediction Accuracy}
The main purpose of this series experiments is to verify the effectiveness of federated learning, by comparing the accuracy of following trained models.
\begin{itemize}
\item Local model:Companies train individual models with $3 \times 15 = 45$ iterations based on the local data set only.
\item Centralized model: Companies upload data sets to a centralized server and the latter trains a model by $3 \times 15 = 45$ iterations based on the consolidated data set.
\item Federated model: Companies join peer-to-peer MPC-enabled model aggregation for $15$ global epochs after training local model with $3$ iterations 
\end{itemize}

Three popular metrics, whose definition could be found in~\cite{b38}, are used to measure prediction accuracy.
\begin{itemize}
\item Recall: The percentages of positive events (i.e., machine faulty cycles) which are predicted correctly.
\item Precision: The percentages of predicted positive events which are positive in truth.
\item Balanced: Overall performance of a model considering both positive and negative classes without worrying about the imbalance of a data set.
\end{itemize}
\begin{table*}[!bth]
\caption{Prediction Accuracy of Differently Trained models} 
\label{AcuracyResult}
\centering
\begin{tabular}{|p{0.5in}|p{0.5in}||p{0.4in}|p{0.4in}|p{0.4in}||p{0.4in}|p{0.4in}|p{0.4in}||p{0.4in}|p{0.4in}|p{0.4in}|}
  \hline

 \textbf{Model Type}& \textbf{Training Method}&  \textbf{Recall Mean}&  \textbf{Recall Highest}&  \textbf{Recall Lowest }&  \textbf{Precision Mean}&  \textbf{Precision  Highest} &  \textbf{Precision Lowest} &  \textbf{Balanced  Mean} &  \textbf{Balanced  Highest} &  \textbf{Balanced  Lowest}\\
\hline

\multirow{11}{*}{SimpleNN} & Local & 0.922 & 1.0 & 0.810 &  0.891 & 1.0 & 0.686 &  0.911  & 0.989   & 0.772 \\
 \cline{2-11} & Centralized & 0.939  & 1.0    & 0.899  &  0.937 & 1.0 & 0.838 &  0.941 & 1.0 & 0.802 \\
 \cline{2-11} & Federated & 0.933  & 1.0 & 0.869  &  0.934 & 1.0 & 0.828 &  0.935 & 1.0 & 0.852 \\
  \hline
  
\multirow{11}{*}{ComplexNN} & Local & 0.913 & 1.0 & 0.803 &  0.921 & 1.0 & 0.7969 &  0.918  & 0.992   & 0.814 \\
 \cline{2-11} & Centralized & 0.952  & 1.0 & 0.881  &  0.945 & 1.0 & 0.828 &  0.949 & 1.0 & 0.850 \\
 \cline{2-11} & Federated & 0.947  & 1.0 & 0.861  &  0.940 & 1.0 & 0.820 &  0.945 & 1.0 & 0.845 \\
  \hline
\end{tabular}
\end{table*}
Four motor data sets (refer to Section~\ref{usecaseSection}) belonging to independent companies were used. We select any three company’s data sets as training data and the remaining one as testing data, in a round-robin manner. Additive and Shamir secret sharing MPC obtained the same experimental results for SimpleNN and ComplexNN models respectively. As shown in Table~\ref{AcuracyResult}, federated learning achieves comparable prediction accuracy to centralized learning, which verify the effectiveness of federated learning for smart manufacturing, without compromising data privacy. Federated learning models outperforms local models, but the advantages for different companies might vary.
\subsection{Communication Cost}
\label{CommunicationResults}
We evaluate the communication cost in terms of the number and size of messages exchanged between all FL parties and/or model aggregation committee members. In traditional peer-to-peer framework, the shared model could be reconstructed only if all peers are working together.  In contrast, in two-phase framework, the shared model could be reconstructed from the collusion among all selected committee members.  We considered the committee with 3 parties, based on the the assumption that no collusion among $>=3$ participants, which is usually acceptable in application where performance trade-off is considered\cite{b39}. The theoretical analysis has been provided in Section~\ref{MPCFLFramework}. Without further specification, SimpleNN model is used; local training takes three iterations (i.e., $t=3$) and global model aggregation takes 15 epochs  (i.e., $e=15$). For Two-Phase MPC-enabled FL framework, three parties are selected as the model aggregation committee (i.e., $m=3$) and the batch size of each round of election is 10 (i.e., $b=10$).  Accordingly to empirical experience, one round is more than sufficient to decide the committee members.
\begin{figure}[!tbh]
\centering
  \includegraphics[width=0.45\textwidth]{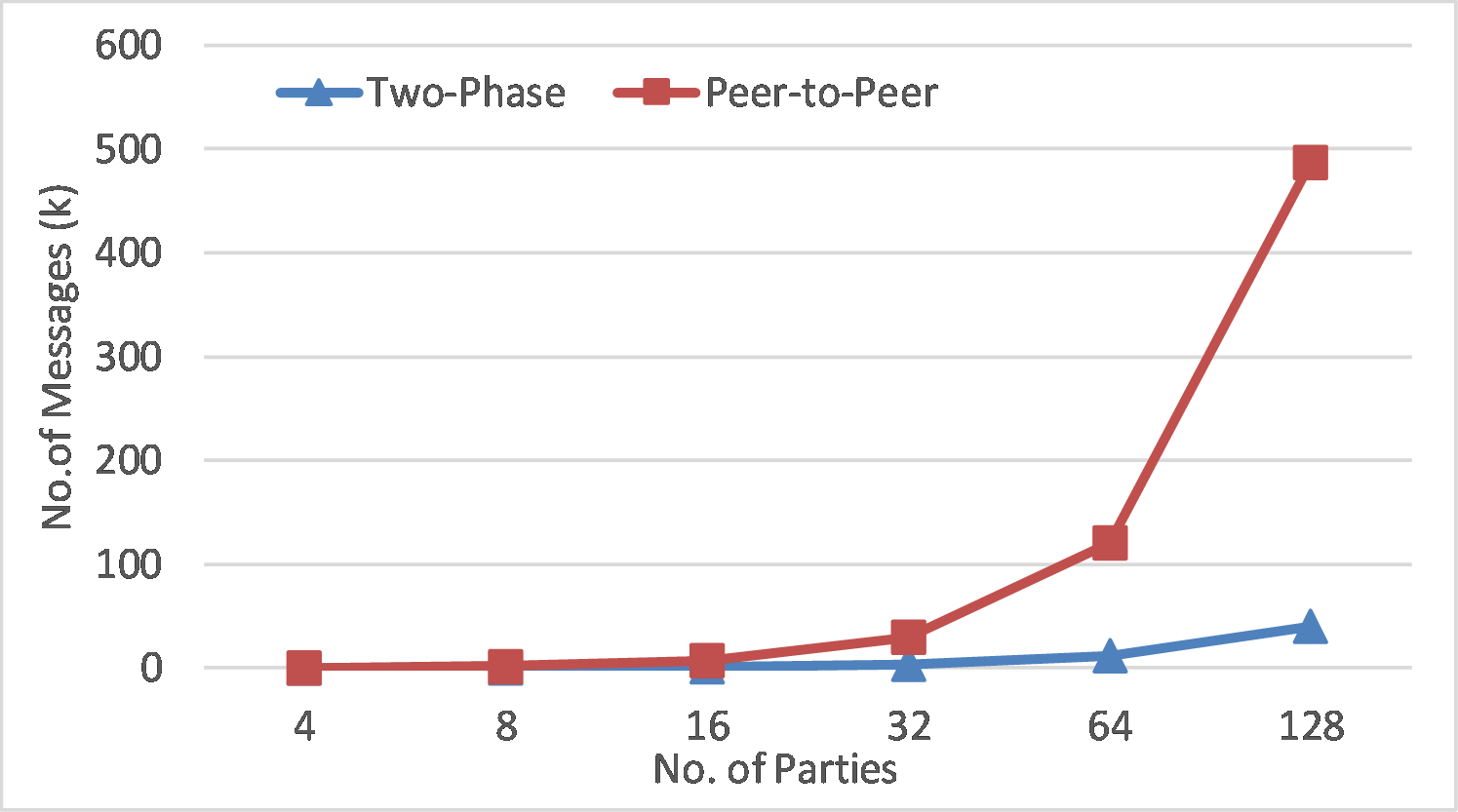}
\caption{Number of Messages VS number of parties on Local-SingleServer}
\label{figure:MsgNum128}       
\end{figure}

\begin{figure}[!tbh]
\centering
  \includegraphics[width=0.45\textwidth]{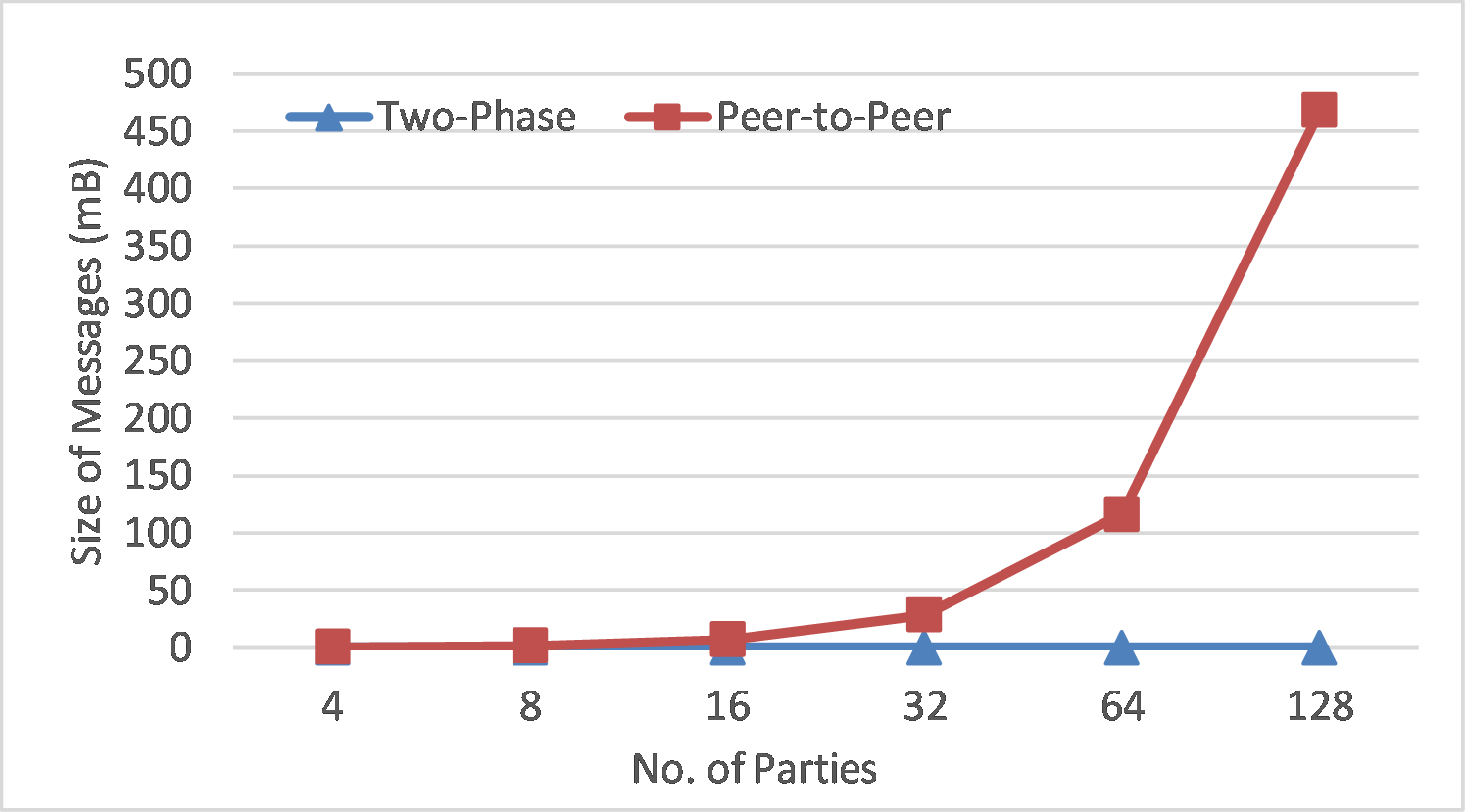}
\caption{Size of Messages VS number of parties on Local-SingleServer}
\label{figure:MsgSize128}       
\end{figure}
\begin{figure}[!tbh]
\centering
  \includegraphics[width=0.45\textwidth]{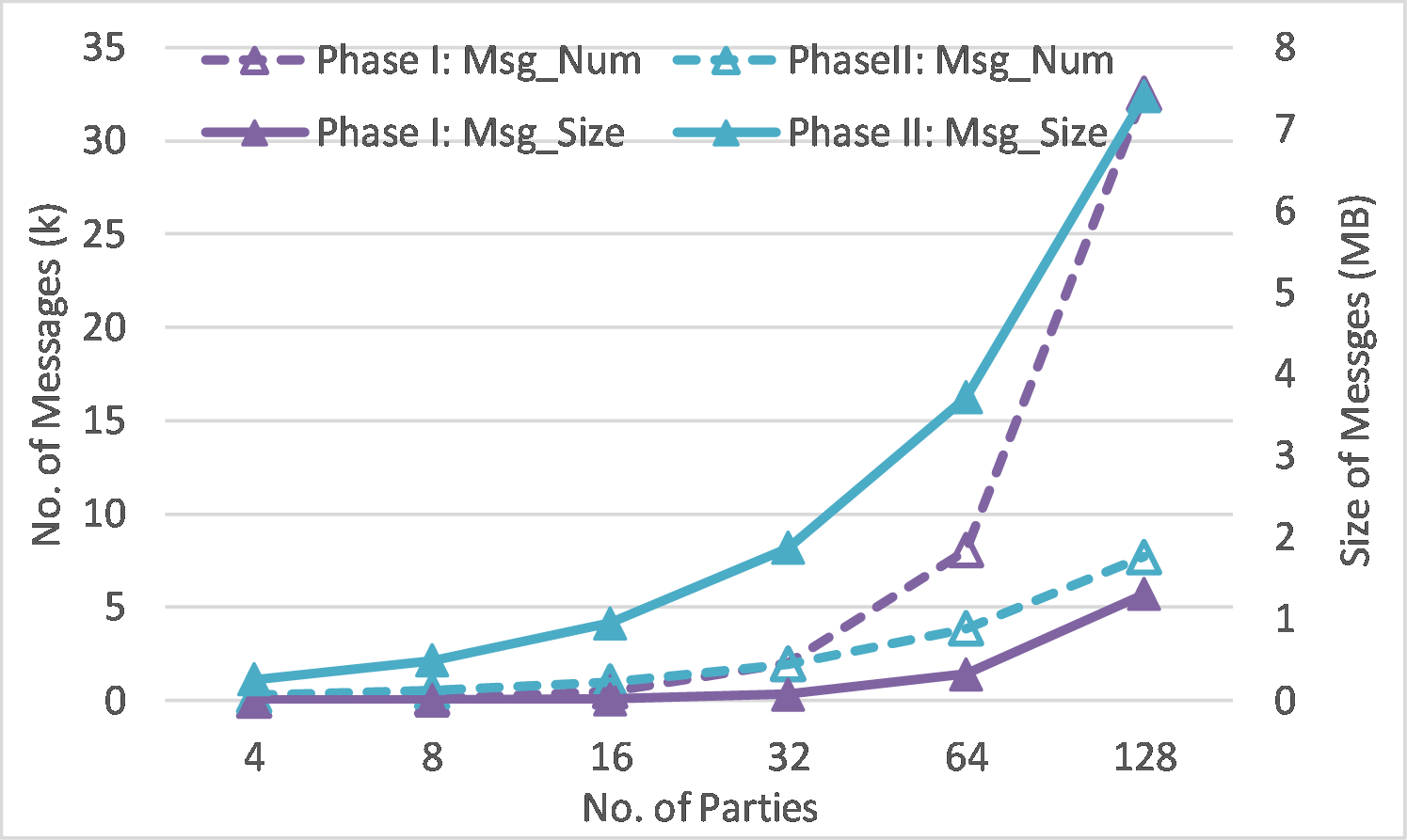}
\caption{Number and Size of messages of two-phase framework on Local-SingleServer}
\label{figure:TwoPhaseMsg128}       
\end{figure}
A series of experiments are conducted on Local-SingleServer environment, by increasing the number of parties (i.e., processes) from 4 to 128. The number and size of messages(depends on the size of the model parameters)  are illustrated on Figure~\ref{figure:MsgNum128} and~\ref{figure:MsgSize128} respectively. The traditional Peer-to-Peer MPC-enabled FL framework has low scalability, as both the number and size of messages increase dramatically (squarely) with respect to the number of parties. As expected in theoretical analysis, the proposed two-phase method dramatically reduces the number and size of messages, and thus improve system scalability.  
More details of Two-Phase framework are further illustrated in Figure~\ref{figure:TwoPhaseMsg128}. When $n<32$, $Phase\ I$ has slightly less messages than $Phase \ II$, but after that, $Phase\ I$ has much more messages than $Phase \ II$. This is because, the number of messages at $Phase\ I$ has $O(n^2)$ complexity (refer to Equation~\ref{msgNumPhaseI}), while the number of messages at $Phase\ II$ has $O(n)$ complexity (refer to Equation~\ref{msgNumPhaseII}). However, since the messages exchanged at $Phase\ I$ are much smaller that those exchanged at $Phase\ II$. (The difference would be even much bigger for ComplexNN model). Hence, it is observed in Figure~\ref{figure:TwoPhaseMsg128} that the size of messages exchanged at $Phase\ I$ is significantly lower than those at $Phase\ II$, although the former may have larger number of messages when a large number of parties join the federated learning. 
\begin{figure}[!tbh]
\centering
  \includegraphics[width=0.45\textwidth,height=0.15\textheight]{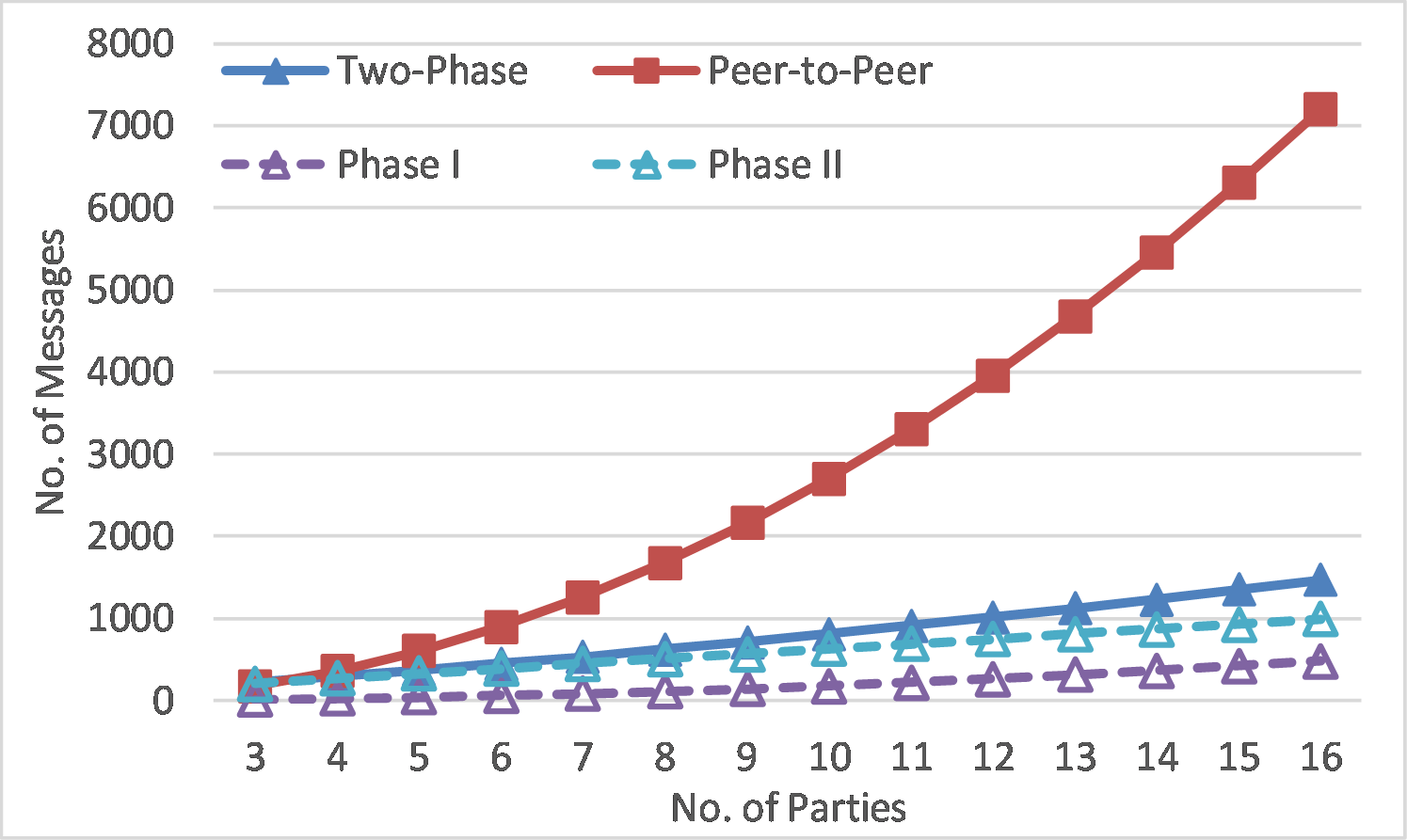}
\caption{Number of Messages VS number of parties on AWS-SameRegion}
\label{figure:MsgNum16}       
\end{figure}
\begin{figure}[!tbh]
\centering
  \includegraphics[width=0.45\textwidth,height=0.15\textheight]{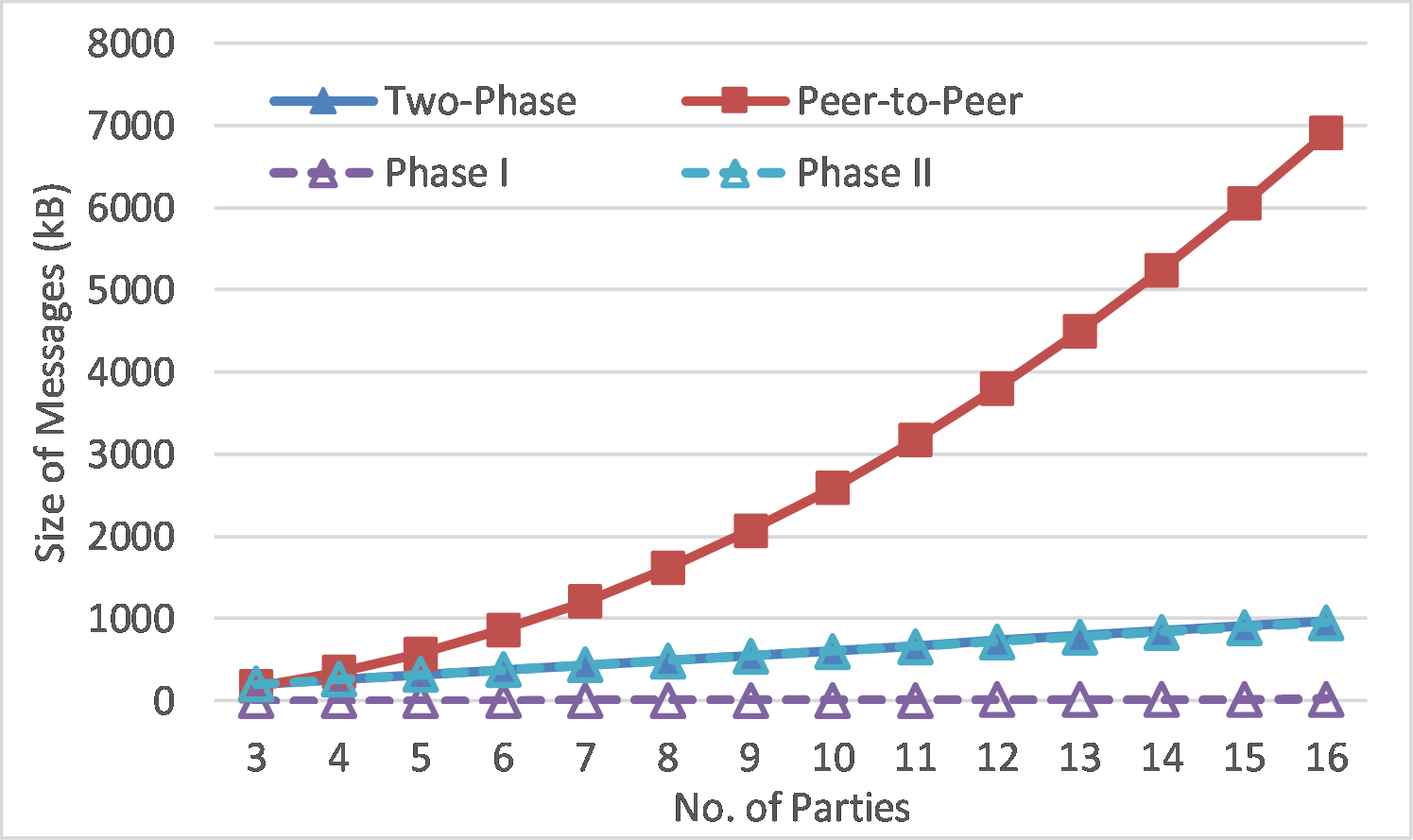}
\caption{Size of Messages VS number of parties on AWS-SameRegion}
\label{figure:MsgSize16}       
\end{figure}
Another two series of experiments are conducted on AWS-SameRegion and AWS-crossRegion environments, by increasing the number of parties from 3 to 16. The number and size of messages measured for AWS-SameRegion and AWS-crossRegion environment are the same. To save space, only AWS-SameRegion scenario is illustrated on Figure~\ref{figure:MsgNum16} and~\ref{figure:MsgSize16}. Similarly to Local-SingleServer execution environment, compared with traditional Peer-to-Peer method, the proposed Two-Phase MPC-enabled FL framework has much lower communication cost and better scalability with respect to the increasing number of parties. Since $n\in [3, 16]$ is considerably low, the number of messages exchanged at $Phase\ I$ (i.e., Equation~\ref{msgNumPhaseI}) is smaller than that at $Phase\ II$ (i.e., Equation~\ref{msgNumPhaseII}). However, the size of messages exchanged at $Phase\ I$ is marginal compared with those at $Phase\ II$  as each message at $Phase\ I$ is a small array of committee votes (in the form of secret shares).   
\subsection{Execution Time}
\begin{figure}[!tbh]
\centering
  \includegraphics[width=0.45\textwidth,height=0.15\textheight]{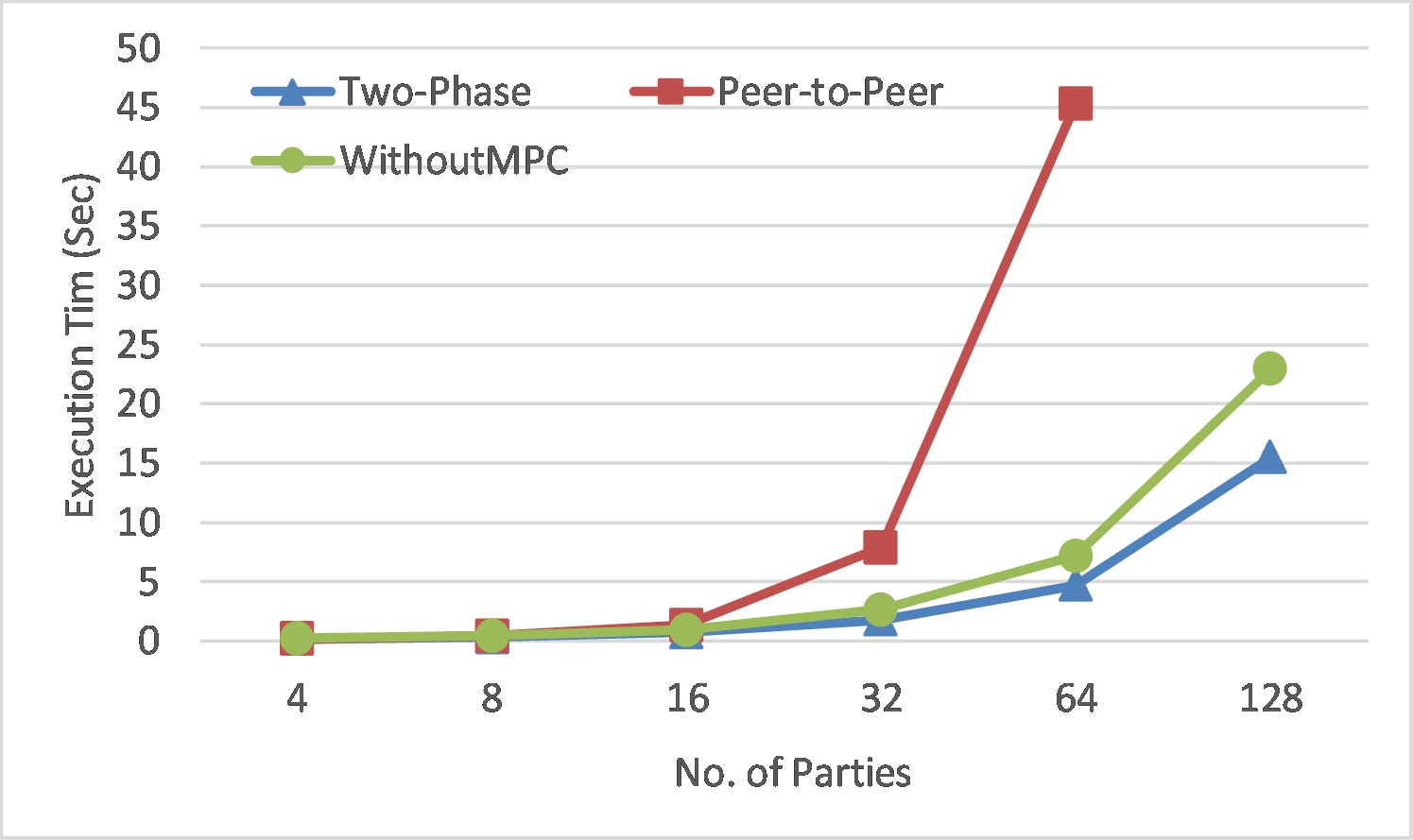}
\caption{Execution time VS number of parties on Local-SingleServer}
\label{figure:Time128}       
\end{figure}
The execution time on local-SingleServer environment is plot on Figure~\ref{figure:Time128}. Traditional peer-to-peer framework is definitely not scalable in terms of the increasing number of parties. Fortunately, by adopting the proposed two-phase MPC election and model aggregation, the system is much more scalable. The execution time is reduced by 25 times, when $n=128$. The execution time of Peer-to-Peer scenario is $390.5sec$ which is out of the scale of Figure~\ref{figure:Time128}.
\begin{figure}[!tbh]
\centering
  \includegraphics[width=0.45\textwidth,height=0.15\textheight]{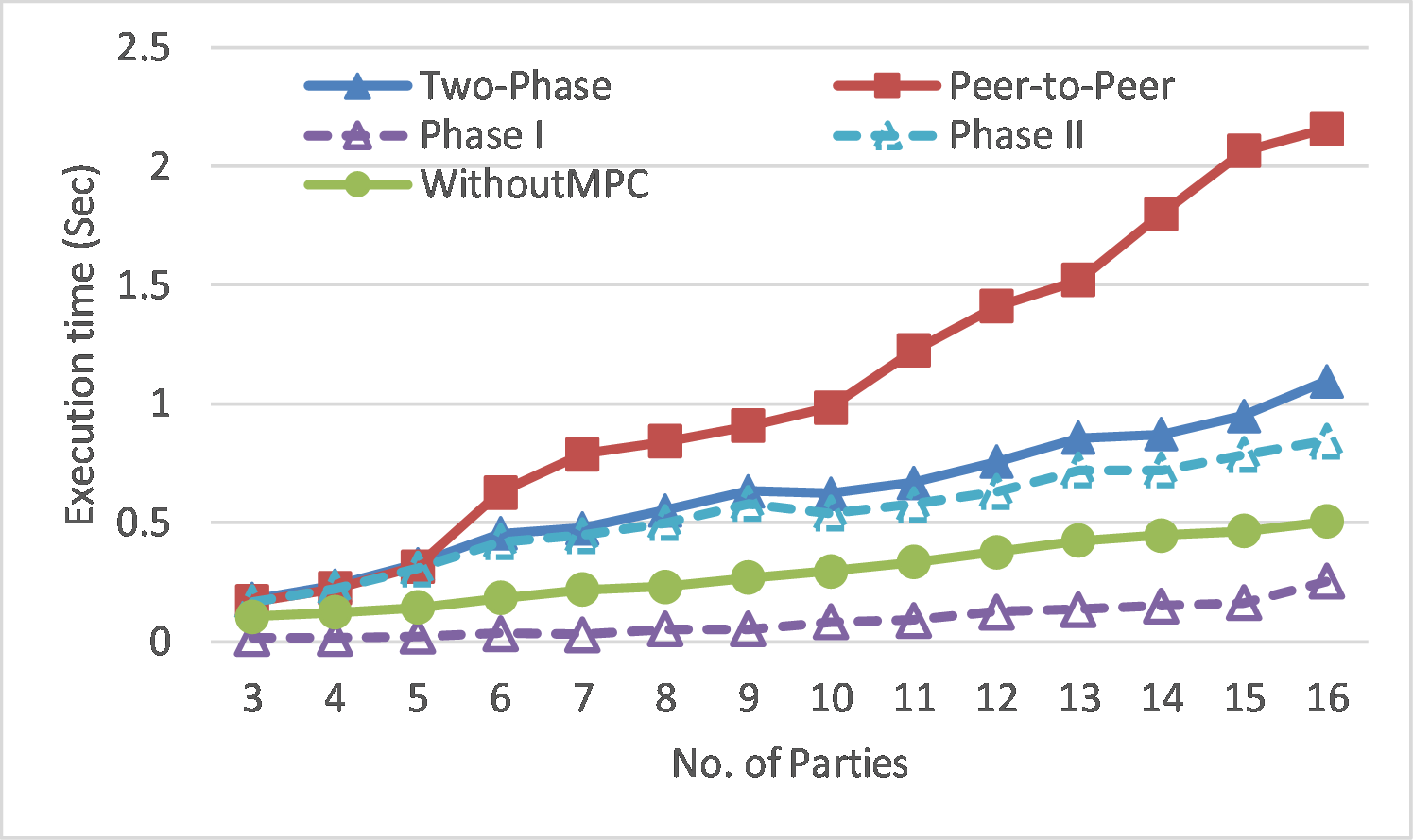}
\caption{Execution time VS number of parties on AWS-SameRegion}
\label{figure:Time16Same}       
\end{figure}
The execution time on AWS-SameRegion environment is plot on Figure~\ref{figure:Time16Same} where 3 to 16 FL parties are executed on individual AWS instances located at Singapore data centre. As expected, the execution time increasing trend of Two-Phase framework is less significant than that of peer-to-peer framework. Execution time of $Phase\ I$ is marginal, as it runs one time only for entire FL training and the size of each messages (i.e., $b=10$) is very small.   
\begin{figure}[!tbh]
\centering
  \includegraphics[width=0.45\textwidth]{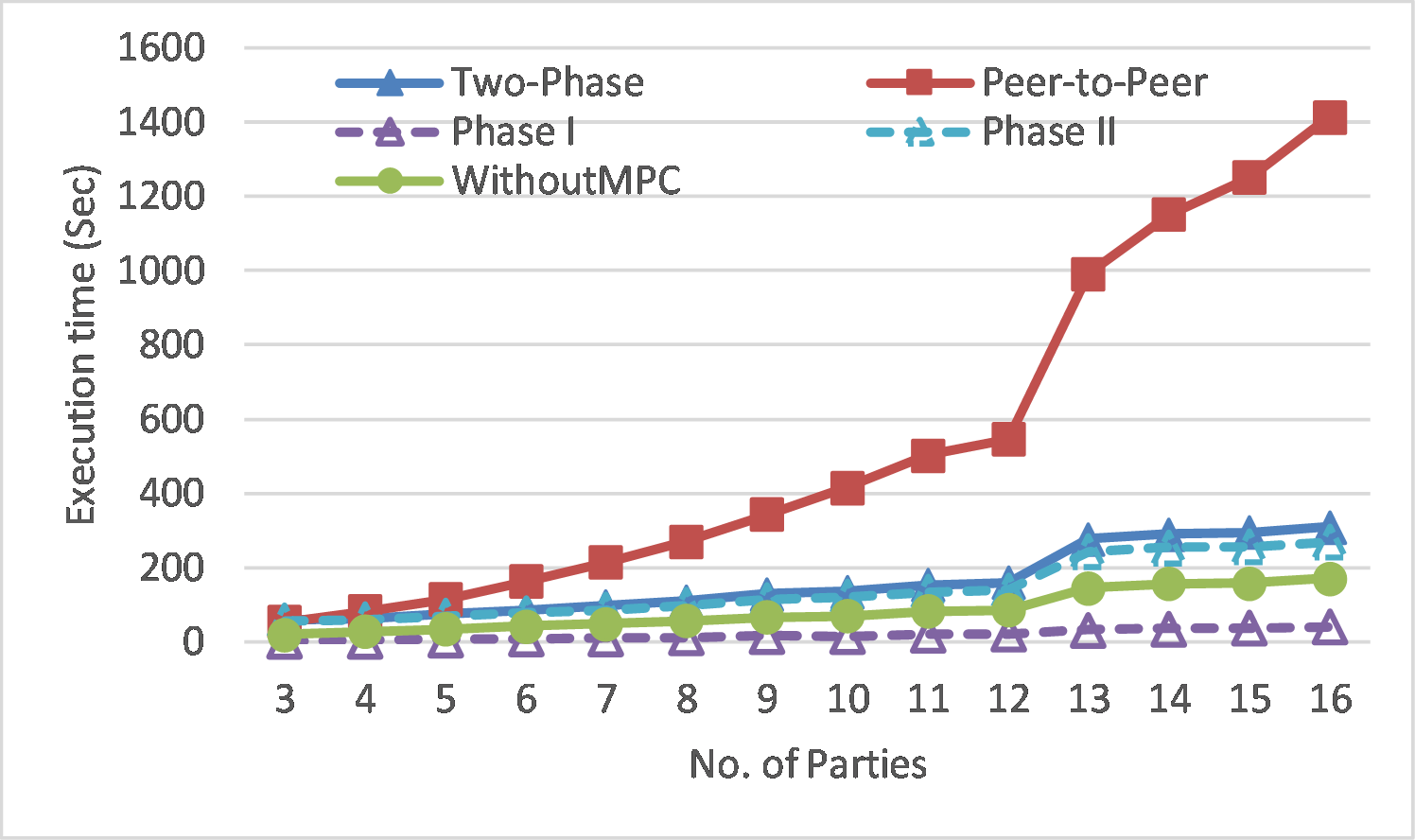}
\caption{Execution time VS number of parties on AWS-CrossRegion}
\label{figure:Time16Cross}       
\end{figure}
The execution time on AWS-CrossRegion environment is plot on Figure~\ref{figure:Time16Cross} where 3 to 16 FL parties are executed on individual AWS instances located at different regions around the world. The execution time increasing trends of all curves are similar to Figure~\ref{figure:Time16Same}. When $n=16$, Two-Phase framework is 1.97 and 4.56 times faster than Peer-to-Peer framework for AWS-SameRegion and AWS-CrossRegion respectively. Due to the network latency and bandwidth limitation across data centres at different regions, the execution time increase hundreds of times. Moreover, erupts are observed when $n$ increase from 12 to 13. This is because we assigned two parties at each region, and the new added parties were located at Sao Paulo (South America) data centre which had significantly worse network condition compared with others.   
In order to evaluate MPC overhead further, execution time of $withoutMPC$ scenario (i.e., all FL parties exchange local models directly in a peer-to-peer manner and then obtain the aggregated model with averaged weights) is also reported in Figure~\ref{figure:Time128},~\ref{figure:Time16Same}, and~\ref{figure:Time16Cross}. When $n\in [3, 16]$, Peer-to-Peer (Two-Phase) MPC-enabled FL framework is around 3.34 and 5.42 times (2.12 and 2.03 times) slower than $withoutMPC$ scenario for AWS-SameRegion and AWS-CrossRegion execution environment respectively. Since Two-Phase MPC-enabled FL framework could improve scalability significantly, it becomes even faster than $withoutMPC$ scenario when $n\geq16$ in local-SingleServer case, as shown in Figure~\ref{figure:Time128} . 
\begin{figure}[!tbh]
\centering
  \includegraphics[width=0.45\textwidth]{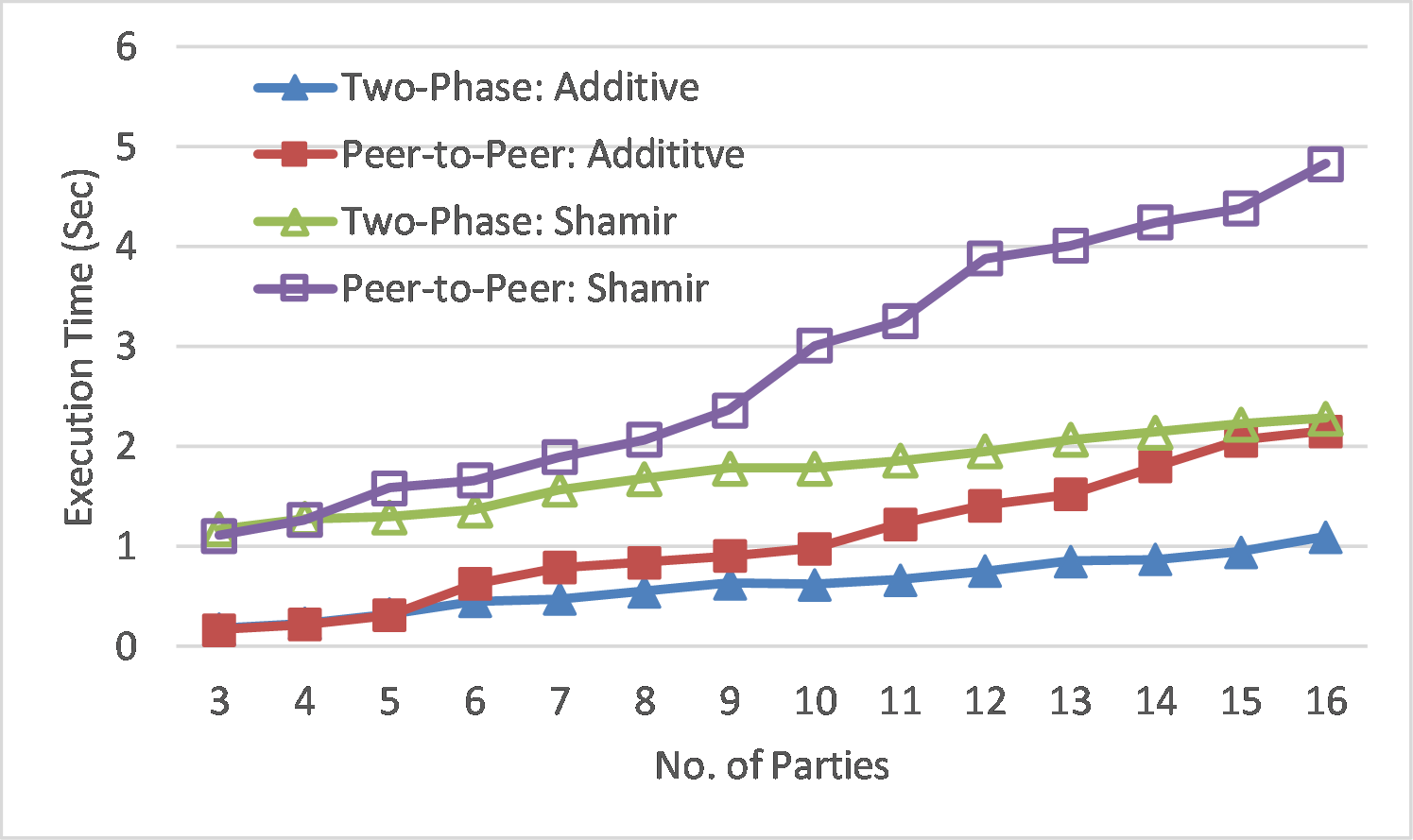}
\caption{Execution Time of Additive and Shamir MPC Protocols on AWS-SameRegion}
\label{figure:AdditiveShamir}       
\end{figure}
\begin{figure}[!tbh]
\centering
  \includegraphics[width=0.45\textwidth]{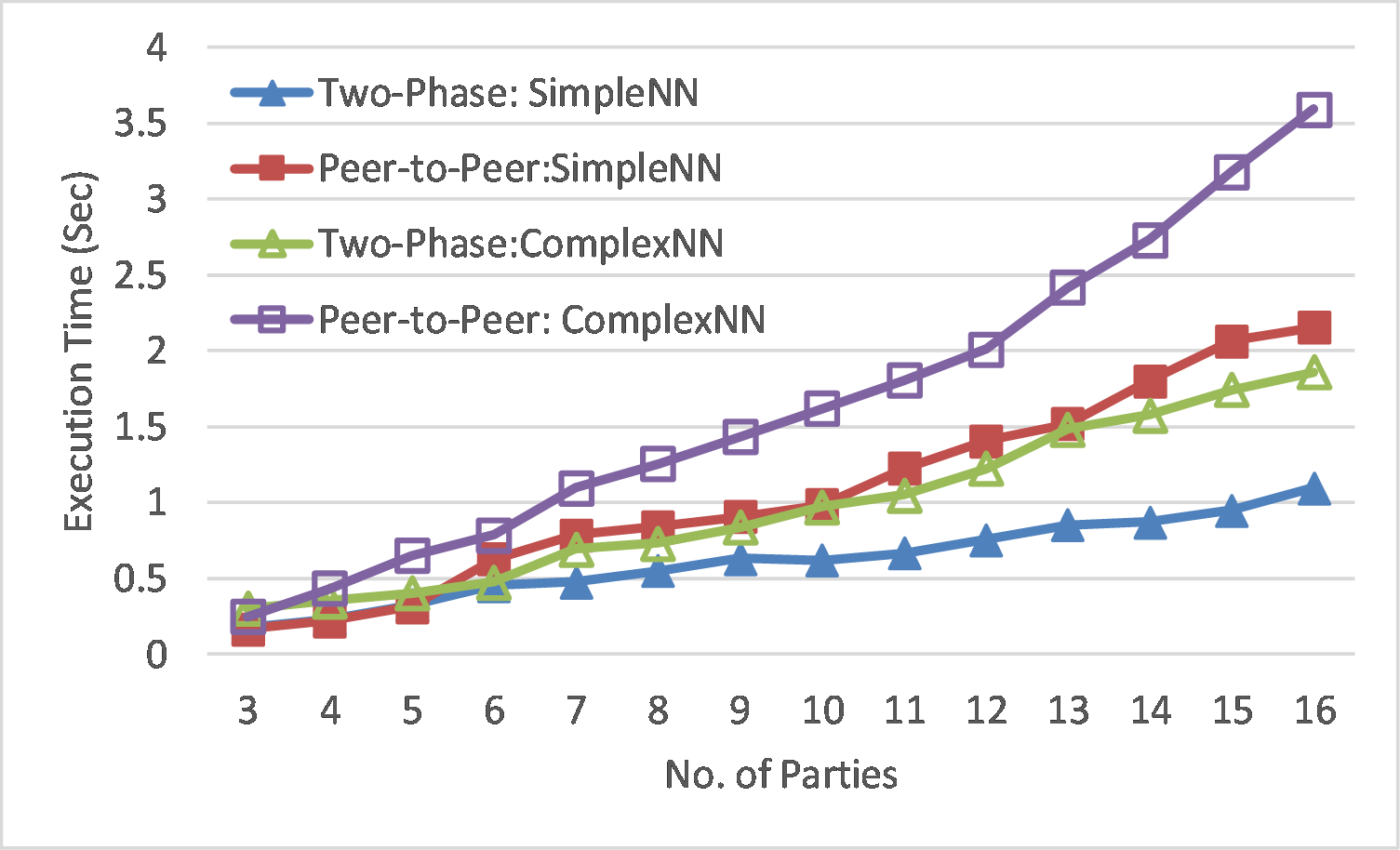}
\caption{Execution Time of SimpleNN and ComplexNN Models on AWS-SameRegion}
\label{figure:SimpleComplex}       
\end{figure}
Figure ~\ref{figure:AdditiveShamir} compares federated learning execution time between Additive and Shamir secret sharing MPC protocols. As mentioned Section~\ref{MPCProtocols}, Shamir protocol requires much heavier computation overhead to reconstruct the polynomial $q(x)$ using Lagrange interpolation. So Shamir MPC protocol needs longer execution time than Additive MPC protocol. The execution time increasing trends with respect to the number of parties are also more significant. When $n=16$, Two-phase MPC-enabled FL framework is 1.97 (Additive) and 2.12 (Shamir respectively) faster than traditional Peer-to-Peer framework. 
Figure ~\ref{figure:SimpleComplex} compares FL execution time between SimpleNN and ComplexNN models. By adding a hidden layer, the model size of ComplexNN increase by 30.5 times. Since large messages are sent efficiently in bulks, the execution time of of FL using ComplexNN is more than two times of that using SimpleNN. However, we can still observe that the increasing trends of those curves using ComplexNN are more significant. When $n=16$, Two-phase MPC-enabled FL framework is 1.97 (SimpleNN) and 1.94 (ComplexNN respectively) faster than traditional Peer-to-Peer framework. 
\section{Conclusion and Future Work}
In this paper we presented a Two-Phase MPC-enabled FL framework. It supports multiple organizations to collectively learn a machine learning model, while keeping private data on-premise. Both Additive and Shamir secret sharing MPC protocols are adopted to aggregate lo- cal models in a privacy-preserving manner. The proposed federated learning framework is further integrated with  an IIoT platform for smart manufacturing. The effectiveness of federated learning has been verified as it can build a better model than that trained on silo data sets and achieves comparable prediction accuracy versus the traditional centralized learning. Compared with the traditional Peer-to-Peer framework, the Two-Phase MPC-enabled FL framework significantly reduce communication cost and improves system scalability. Depending on the number of parties, it achieves 2 to 25 times of speed-up on execution time of federated learning for both simple and complex neural network models. 
In the future, we will extend our work to include various application domains to exploit the advantages of FL in practice. Besides horizontal FL, vertical FL and transfer learning~\cite{b34} will also be investigated to enable collaborations across multiple domains. Further, we will continue our research on system development for performance and scalability enhancement by considering large number of parties on AWS-SameRegion and AWS-CrossRegion with large datasets, as well as on privacy-preserving and cyber security techniques to deal with malicious users who negatively affect federated learning effectiveness in a trustless environment~\cite{b9}.
\section*{Acknowledgment}
This research was supported by Grant No:A1918g0063, Trusted Data Vault(Phase 1), RIE 2020 Advanced Manufacturing And Engineering(AME) Domain's Core Funds-SERC Strategic Funds.

\end{document}